\documentclass[12pt]{article}
\def\beq{\begin{eqnarray}}    
\def\eeq{\end{eqnarray}}      

\def\ln{\,\mbox{ln}\,}                  
\def\Box{\nabla^2}                      
\def\diag{\,\mbox{diag}\,}              

\def\al{\alpha}
\def\be{\beta}

\def\ga{\gamma}
\def\de{\delta}

\def\la{\lambda}
\def\na{\nabla}
\def\pa{\partial}

\def\si{\sigma}

\def\ph{\varphi}

\def\th{\theta}

\def\Ga{\Gamma}
\def\De{\Delta}

\def\Om{\Omega}

\begin{document}

\begin{center}

{\large \sc
On Useful Conformal Tranformations In General Relativity}

\vskip 4mm

%
{\bf
D.F. Carneiro, \
E.A. Freitas,  \  B. Gon\c{c}alves,  \
A.G. de Lima, \ and \ I.L. Shapiro}


\vskip 5mm

{\small\sl
Departamento de F\'{\i}sica, ICE,
Universidade Federal de Juiz de Fora}
\vskip 1mm

{\small\sl
Juiz de Fora, CEP: 36036-330, MG,  Brazil}

\end{center}

\vskip 5mm

\begin{quotation}

\noindent
\textit{Abstract.}
\quad
{Local conformal transformations are known as a useful
tool in various applications of the gravitational theory,
especially in cosmology. We describe some new aspects
of these transformations, in particular using them for
derivation of Einstein equations for the cosmological
and Schwarzschild metrics. Furthermore, the conformal
transformation is applied for the dimensional reduction
of the Gauss-Bonnet topological invariant in $d=4$ to
the spaces of lower dimensions.}
\end{quotation}
\vskip 12mm

\section{Introduction}

General Relativity (GR) is a successful  relativistic
theory of gravitation. According to the existing data,
its predictions fit nicely to the most of the tests and
the limits of its validity are likely to emerge only in
the extreme situations when one has to account for, e.g.
quantum effects \cite{LL-2,Weinberg72,MTW,wald}. The
quantum theory
of matter fields in curved space-time has achieved many
solid results (see e.g. \cite{birdav,fulling,book}, and
may be viewed as a source of important applications
in cosmology and black hole physics. Some of these
achievements are related to the so-called conformal
anomaly, which is nothing but the quantum violation of
the local conformal symmetry which is imposed at the
classical level. In order to illustrate this statement
it is sufficient to mention that this anomaly is
related to such relevant phenomenon as the black hole
evaporation (Hawking effect). The anomaly is also related
to the derivation of the low-energy effective
action of (super)string - the main candidate to be the
unified theory of quantum gravity and other fundamental
interactions.

The importance of local conformal symmetry makes natural
the idea to include it into the basic course of GR.
However, the standard point of view is that such extension
of the program is difficult because of the shortage
of time. In this article we shall suggest a new possibility
that including the conformal transformation into the
program may be compatible with certain saving of time and
efforts due to the relevant simplification in the derivation
of Einstein equations for the metrics related to the most
fundamental applications of GR such as cosmological and
Schwarzschild solutions. Furthermore, we shall use the
conformal transformation to investigate a problem which
goes slightly beyond the standard course of GR, that
is establish the relations between the topological
invariants in four and two dimensional spaces via the
restricted dimensional reduction. It is supposed that this
relation will be useful for those willing to learn more
advanced aspects of the gravitational theory.

The paper is organized as follows. In section 2 we
shall present a brief review of the main properties
of the curvature tensor (see, e.g.
\cite{petrov,Weinberg72,DNF} for the introduction).
In particular, we shall
prove a simple factorization theorem describing
important properties of curvature for the metric
of the special factorized form. In section 3 the
list of formulas concerning the local conformal
transformation in the space of an arbitrary dimension
is presented. For the sake of generality we shall
include the transformations not only for the curvature
tensors but also for the quadratic invariants built
from this tensor and its derivatives. In section 4
the derivation of the Einstein equations for the
cosmological and spherically symmetric metrics is
presented. Section 5 is devoted to establishing the
relation between the topological invariants in
dimensions 4, 2  and 3 via the factorization theorem
from the section 2 and conformal transformations.
Finally, in the last section we draw our conclusions.

\section{Brief review of curvature tensor}

Let us start with the properties of the curvature tensor
which will be used below. A comprehensive introduction
can be found in the monographs on GR and Differential
Geometry \cite{petrov,Weinberg72,wald,MTW,DNF}.

The action of covariant derivative on the tensor
$\,T^{\mu_1 \mu_2 ... \mu_k}\,_{\nu_1\nu_2 ... \nu_l}\,$
is defined by the relation
\beq
\nabla_{\alpha} T^{\mu_1 \mu_2 ... }\,_{\nu_1\nu_2 ...}
&=& \pa_{\alpha}
T^{\mu_1 \mu_2 ... }\,_{\nu_1\nu_2 ... }
+ \Gamma^{\mu_1}_{\lambda \alpha}
T^{ \la\mu_2 ...}\,_{\nu_1\nu_2 ...}
+ \Gamma^{\mu_2}_{\lambda \alpha}
T^{\mu_1\la  ... \mu_k}\,_{\nu_1\nu_2 ... \nu_l}+...
\nonumber
\\
&-& \Gamma^{\lambda}_{\nu_1 \alpha}
T^{\mu_1 \mu_2 ... \mu_k}\,_{\la\nu_2 ... \nu_l}
- \Gamma^{\lambda}_{\nu_2 \alpha}
T^{\mu_1 \mu_2 ... \mu_k}\,_{\nu_1\la ... \nu_l}\,-\,...\,\,,
\label{man 11}
\eeq
providing tensor transformation rule under the general
coordinate transformations.
The coefficients $\,\Gamma^{\lambda}_{\alpha \beta}\,$ are
defined according to
\beq
\Gamma^{\lambda}_{\mu\nu}
= \frac{1}{2}\,g^{\la\tau} \left(\pa_{\mu} g_{\tau\nu}
+ \pa_{\nu}g_{\tau\mu} - \pa_{\tau} g_{\mu\nu}\right)\,,
\label{man 16}
\eeq
such that the covariant derivative $\,\na_\al\,$ satisfies
the metricity condition $\,\na_\al g_{\mu\nu}=0\,$
and also is torsionless
$\,\Gamma^{\lambda}_{\mu\nu}=\Gamma^{\lambda}_{\nu\mu}$.

Consider the commutator of the
two covariant derivatives over the vector $\,T^\al$.
By direct calculations we find
\beq
\left[\na_\mu\,,\,\na_\nu\right]\,T^\al
\,=\, \na_\mu\na_\nu \,T^\al - \na_\nu\na_\mu \,T^\al
\,=\,-\,T^\la\cdot R^\al_{\, \cdot\,\la \,\nu\mu}\,,
\label{curve 7}
\eeq
where
\beq
R^\al_{\, \cdot\,\la\nu\mu}\,=\, \pa_\nu \Ga^\al_{\la\mu}
- \pa_\mu \Ga^\al_{\la\nu}
+\Ga^\tau_{\la \mu}\,\Ga^\al_{\tau \nu}
-\Ga^\tau_{\la \nu}\,\Ga^\al_{\tau \mu}
\label{curve 10}
\eeq
is called (Riemann) curvature tensor.

The relevance of the Riemann tensor is due to its
universality. One can easily show that the commutator
of the two covariant derivatives acting on any tensor
is a linear combination of the curvature tensors, e.g.
\beq
\left[\na_\mu\,,\,\na_\nu\right]\,W^\al_\be
= R^\la_{\,\cdot\, \be\nu \mu}\,W^\al_\la
\,-\, R^\al_{\, \cdot\,\la\nu\mu}\,W^\la_\be \,\,.
\label{curve 12}
\eeq

The curvature tensor has following algebraic symmetries,
which can be better seen on the completely covariant
version $\,R_{\al\be\mu\nu}
\,=\, g_{\al\ga}\,R^\ga_{\,\cdot\,\be\mu\nu}\,$:
\beq
R_{\al\be\mu\nu}\,=\,-\,R_{\al\be\nu\mu}
\,=\,R_{\be\al\nu\mu}\,=\,R_{\mu\nu\al\be}\,,
\label{curve 18}
\eeq
where symmetry between the pairs of the indices
supplements the antisymmetry in the indices of
each pair, and
\beq
R_{\al\be\mu\nu}+R_{\al\nu\be\mu}+R_{\al\mu\nu\be}\,=\,0\,.
\label{curve 26}
\eeq

It proves useful to define the Ricci tensor
as the contraction of the Riemann curvature
\beq
R_{\mu\al}\,=\,R^\be_{\,\cdot\,\al\be\mu}
\,=\,g^{\nu\be}\,R_{\nu\al\be\mu}\,.
\label{curve 29}
\eeq
The Ricci tensor is symmetric $\,R_{\mu\nu}\,=\,R_{\nu\mu}$.
Further contraction produces the scalar curvature
$\,R=R^\al_\al$\,.

The algebraic properties of the Riemann, Ricci tensors
and of the scalar curvature are
the same in the Riemann or pseudo-Riemann spaces of any
dimension. However, in the spaces of lower dimensions
the curvature tensors are more restricted than in the
general $D$-dimensional case. The extreme case is the
$\,D=1\,$ space,
where all curvatures vanish identically.

In order to understand the possible role of $\,D\,$ better,
one has to calculate the number of independent components
of the tensors $\,R_{\mu\nu\al\be}\,$ and $\,R_{\mu\nu}$.
Due to the algebraic symmetries (\ref{curve 18}), the
curvature tensor with 3 equal indices vanish. Therefore,
we have only three possible situations:

{\it 1)}.\
Two couples of equal indices, that is the construction like
$\,R_{0101}$. The combinatorial calculus tells us that the
number of distinct combinations of this type in a
$D$-dimensional space is $\,n_1=C_D^2=\frac{D(D-1)}{2}$.

{\it 2)}.\
One couple of equal indices plus two distinct indices,
that is the construction like $\,R_{0102}$. The
number of distinct combinations here is
$\,n_2=D\,C_{D-1}^2=\frac{D(D-1)(D-2)}{2}\,$.

{\it 3)}.\
Four distinct indices, that is the construction like
$\,R_{0123}$. In this case we have
$$
n_3=\frac{2D(D-1)(D-2)(D-3)}{4!}\,.
$$

Summing up the three expressions, we obtain the number
of independent components of the Riemann tensor in
$\,D$-dimensional space.
\beq
N_D\,=\,n_1+n_2+n_3\,=\,\frac{D^2(D^2-1)}{12}\,.
\label{number 1}
\eeq

Now we are in a position to consider the particular cases
of lower dimensions. Let us start from the $\,D=2\,$ case,
where only the option {\it 1)} takes place.
$\,N_2=1\,$ and we see that the number of independent
components of the Riemann tensor equals to the one of the
scalar curvature. Since both of them are linear combinations
of the partial derivatives of the metric, they must be
proportional. Taking the algebraic symmetries
into account, we obtain
$$
R_{\mu\nu\al\be}\,=\,t\,R\,\left(g_{\mu\al}\,g_{\nu\be}
- g_{\mu\be}\,g_{\nu\al}\right)\,,
$$
where $\,t\,$ is an unknown coefficient. Making a
contraction, we arrive at
\beq
R_{\mu\al}\,=\,t\,R\,g_{\nu\be}\qquad
\mbox{and}\qquad R\,=\,2\,t\,R\,.
\label{number 2}
\eeq
Hence $t=1/2$ and we obtain well-known relations
\beq
R_{\mu\nu\al\be}\,=\,\frac12\,
R\,\left(g_{\mu\al}\,g_{\nu\be}-g_{\mu\be}\,g_{\nu\al}\right)
\,,\qquad
R_{\mu\al}\,=\,\frac12\,R\,g_{\nu\be}\,.
\label{number 3}
\eeq

For $\,D=3\,$ space we have, according to (\ref{number 1}),
six independent components of both Riemann and Ricci tensors.
The Riemann tensor is a linear combination of the
scalar curvature and the Ricci tensor $\,R_{\mu\nu}$.
Taking into account the algebraic symmetries and
making contractions we obtain the relation which always
holds for $D=3$ but not necessary for $D\geq 3$
\beq
R_{\mu\nu\al\be}
\,=\,\left(R_{\mu\al}\,g_{\nu\be}-R_{\mu\be}\,g_{\nu\al}
+R_{\nu\be}g_{\mu\al}-R_{\nu\al}g_{\mu\be}\right)
-\frac12R\left(g_{\mu\al}g_{\nu\be}
-g_{\mu\be}g_{\nu\al}\right).
\label{number 5}
\eeq

Obviously, the relation (\ref{number 5}) means that there
is some part of the Riemann tensor which automatically
vanish in $\,D=3\,$ but may be non-zero for $D>3$.
The corresponding object is called Weyl tensor and is
defined as follows
\beq
C_{\mu\nu\al\be}\,=\,R_{\mu\nu\al\be}
\,-\,\frac{1}{D-2}\,
\left(R_{\mu\al}\,g_{\nu\be}-R_{\mu\be}\,g_{\nu\al}
+R_{\nu\be}\,g_{\mu\al}-R_{\nu\al}\,g_{\mu\be}\right)
\nonumber
\\
\,+\,\frac{1}{(D-1)(D-2)}\,R\,\left(g_{\mu\al}\,g_{\nu\be}
-g_{\mu\be}\,g_{\nu\al}\right)\,.
\label{number 6}
\eeq
The Weyl tensor has, by construction,  the same algebraic
symmetries as the Riemann tensor and also two additional
properties
\beq
g^{\mu\al}\,C_{\mu\nu\al\be}(D)\,=\,0\,,\qquad
C_{\mu\nu\al\be}(D=3)\,\equiv\,0\,.
\label{Weyl}
\eeq
Another important feature of the Weyl tensor will be discussed
in the next section.

The last sort of curvature which we have to
define is called Einstein tensor
\beq
G_{\mu\nu}\,=\,R_{\mu\nu}
\,-\,\frac{1}{2}\,R\,g_{\mu\nu}\,.
\label{number 7}
\eeq

The Einstein tensor is constructed, exactly as Riemann,
Ricci, Weyl tensors and scalar curvature, from the metric
and its derivatives. All these tensors are homogeneous
functions of the second order in the derivatives.
The importance of the Einstein tensor is due to the fact
that the Einstein equations, which define the gravitational
fields in General Relativity, have the form
\beq
G_{\mu\nu}\,=\,8\pi G\,T_{\mu\nu}\,,
\label{number 8}
\eeq
where $\,G\,$ is the Newton constant and $\,T_{\mu\nu}\,$
is a source term which is called Energy-Momentum Tensor
(or stress tensor). In the next sections we shall develop
the efficient method to calculate the Einstein tensor
and, if necessary, also Riemann, Ricci and Weyl tensors,
for the particular metrics of special physical interest.

The metric which provides $\,R_{\mu\nu\al\be}=0\,$
are necessary flat. However, there are situations
when $\,R_{\mu\nu\al\be}\neq 0$, but some of the reduced
versions of curvature vanish. One can distinguish
Ricci flat or Weyl flat metrics, which are not flat
in the proper sense but provide the vanishing Ricci
or Weyl tensors, correspondingly.
As we already know, any $\,D=3\,$ space is Weyl flat,
for example.

In many problems of GR we do not need an arbitrary
metric but instead just a particular form with a smaller
number of independent components than the general one.
Sometimes it is possible to reduce the derivation
of the curvature tensors for these particular
metrics to the one in a lower dimensional case.
Let us now formulate the theorem about the factorized
metrics which will prove very useful in what follows.
\vskip 1mm

{\bf Theorem.}\quad
Consider the $\,D\,$-dimensional Riemann or pseudo-Riemann
manifold with the coordinates
$\,x^\mu=\left(x^a,\,x^i\right)$, where
$\,a,b,c,...=1,2,...,n\,$ and $\,i,j,k,...=n+1,n+2,...,D$.
Let us assume that the metric is factorized
\beq
g_{\mu\nu}\,=\,\left\|
\begin{array}{cc}
g_{ab}(x^a)  &  0           \\
0            &  g_{ij}(x^i) \end{array}
\right\|\,.
\label{number 9}
\eeq
For this metric the elements with the mixed
indices $\,g_{ai}\,$ are zero, and only the blocks with
$\,g_{ab}\,$ and $\,g_{ij}\,$ do not vanish.
Moreover, the metric elements in each block
depend only on their ``own'' coordinates
$\,g_{ab}=g_{ab}(x^a)\,$ and $\,g_{ij}=g_{ij}(x^i)$.
The statement is that, for the metric (\ref{number 9}),
the Riemann tensor is also factorized. This means that
all its components with the mixed indices do vanish.
\vskip 2mm

{\bf Proof.}\quad
First of all, it clear that the inverse metric will
also have the factorized block structure
\beq
g^{\mu\nu}\,=\,\left\|
\begin{array}{cc}
g^{ab}(x^a)  &  0           \\
0            &  g^{ij}(x^i) \end{array}
\right\|\,.
\label{number 10}
\eeq
Consider the Christoffel symbol with the mixed
indices, e.g.
\beq
\Ga^a_{ib}\,=\,g^{ac}\left( \pa_i g_{bc}
+\pa_b g_{ic}-\pa_c g_{ib} \right)\,.
\label{number 11}
\eeq
Taking into account the relations $\,g_{ic}=0\,$ and
$\,\pa_i g_{bc}=0$, we obtain $\,\Ga^a_{ib}=0$. In the
same way one can prove that other mixed components
vanish
\beq
\Ga^i_{jb}\,=\,\Ga^i_{ab}\,=\,\Ga^a_{ij}\,=\,0\,.
\label{number 12}
\eeq
Thus, only the ``pure'' components of the Christoffel
symbol  $\,\Ga^a_{bc}(x^a)\,$ and $\,\Ga^k_{ij}(x^i)\,$
can be non-zero.

Let us now consider the Riemann tensor with the
mixed components. According to the formula (\ref{curve 10})
it is zero because {\it i)} the ``mixed'' derivatives
(e.g. $\,\pa_i\Ga^a_{bc}$) equal zero; \
{\it ii)} the $\,\Ga \Ga$-type terms with the mixed
indices can not be constructed from the $\,\Ga$-symbols
with the ``pure'' indices.

The last observation is that one can easily extend the
factorization theorem for the case of Ricci tensor. The
scalar curvature is given by the sum
$\,\,R(g_{\mu\nu}) = R(g_{ab}) + R(g_{ij})\,$ and the
form of factorization of the Weyl tensor can be easily
obtained from its definition (\ref{number 6}).

\section{Local conformal transformations}

Many properties of the physically interesting metrics
become more apparent and easier to prove if we use the
so called local conformal transformation
\beq
g_{\mu\nu}(x)\,=\,{\bar g}_{\mu\nu}(x)\,e^{2\si(x)}\,.
\label{c 1}
\eeq
(\ref{c 1}) is not the coordinate transformation, it
can be seen as an alternative form of parametrizing the
metric via the new variables $\,{\bar g}_{\mu\nu}(x)\,$
and $\,\si(x)$. Our purpose in this section will be
to derive the relevant quantities like curvatures in
this new parametrization.
Since some of the formulas below are bulky, we shall
use the condensed notations
$\,(\na \si)^2=g^{\mu\nu}\na_\mu \si \na_\nu \si\,$
and $(\na^2 \si)=g^{\mu\nu}\na_\mu \na_\nu \si$.

The transformation laws for the inverse metric and
for the metric determinant have the form
\beq
g^{\mu\nu}={\bar g}^{\mu\nu}\,e^{-2\sigma}\,,
\qquad g={\bar g}\,e^{2D\sigma} \,.
\label{c 2}
\eeq

For the Christoffel symbols we have
\beq
{\Ga}^\la_{\al\be} = {\bar \Ga}^\la_{\al\be}
+ \de{\Ga}^\la_{\al\be}
\,,\qquad
\de {\Ga}^\la_{\al\be}
\,=\,\de^\la_\al \,{\bar \na}_\be\si
\,+\, \de^\la_\be \,{\bar \na}_\al\si
\,-\, {\bar g}_{\al\be} \,{\bar \na}^\la\si\,,
\label{c 3}
\eeq
where $({\bar \na}_\al\si)=\pa_\al\si$. Here and below
all quantities with bars (e.g. ${\bar \na}_\al$
or ${\bar R}_{\al\be}$ correspond to the metric
$\,\,{\bar g}_{\mu\nu}$.
Let us remark that the quantity $\de{\Ga}^\la_{\al\be}$
is a tensor, while ${\Ga}^\la_{\al\be}$ and
${\bar \Ga}^\la_{\al\be}$ are not.

In order to derive the transformation for the Riemann
tensor one can start from the simple relation
\beq
R^\al_{\,\cdot\,\be\mu\nu}
&=& {\bar R}^\al_{\,\cdot\,\be\mu\nu}
\,+\,{\bar \na}_\mu \de\Ga^\al_{\be\nu}
- {\bar \na}_\nu \de\Ga^\al_{\be\mu}
+ \de\Ga^\tau_{\be \nu}\,\de\Ga^\al_{\tau \mu}
- \de\Ga^\tau_{\be \mu}\,\de\Ga^\al_{\tau \nu}\,.
\label{bunch 1}
\eeq
Starting from (\ref{c 1}), after some algebra we obtain
\beq
R^\al_{\,\cdot\,\be\mu\nu}
&=& {\bar R}^\al_{\,\cdot\,\be\mu\nu}
+\de^\al_\nu\,({\bar \na}_\mu{\bar \na}_\be\si)
-\de^\al_\mu({\bar \na}_\nu{\bar \na}_\be\si)
+{\bar g}_{\mu\be}({\bar \na}_\nu{\bar \na}^\al\si)
\nonumber
\\
&-& {\bar g}_{\nu\be}({\bar \na}_\mu {\bar \na}^\al\si)
+\de^\al_\nu {\bar g}_{\mu\be}({\bar \na} \si)^2
-\de^\al_\mu {\bar g}_{\nu\be}({\bar \na} \si)^2
+ \de^\al_\mu({\bar \na}_\nu\si)({\bar \na}_\be\si)
\nonumber
\\
&-& \de^\al_\nu({\bar \na}_\mu\si)({\bar \na}_\be\si)
+ {\bar g}_{\nu\be}({\bar \na}_\mu\si)({\bar \na}^\al\si)
- {\bar g}_{\mu\be}({\bar \na}_\nu\si)({\bar \na}^\al\si)
\label{c Riemann}
\eeq
and furthermore
\beq
R_{\mu\nu}\,=\,{\bar R}_{\mu\nu}
- {\bar g}_{\mu\nu}({\bar \na}^2 \si)
+ (D-2)\left[({\bar \na}_\mu\si)({\bar \na}_\nu\si)
- ({\bar \na}_\mu {\bar \na}_\nu\si)
-  {\bar g}_{\mu\nu}({\bar \na}\si)^2\right]
\label{c Ricci}
\eeq
and
\beq
R\,=\,e^{-2\sigma}\,\Big[\,{\bar R}
- 2(D-1)({\bar \na}^2 \sigma)
- (D-1)(D-2)({\bar \na} \sigma)^2\,\Big]\,.
\label{c scalar}
\eeq

It is easy to see, using (\ref{c Riemann}),
(\ref{c Ricci}) and (\ref{c scalar}),
that the Weyl tensor (\ref{number 6}) transforms in a most
simple possible way
\beq
C^{\al}_{\,\cdot\,\be\mu\nu}
\,=\,{\bar C}^\al_{\,\cdot\,\be\mu\nu}\,,
\qquad {\rm i.e.}\qquad
C_{\al\be\mu\nu}\,=\,e^{2\si}\,{\bar C}_{\al\be\mu\nu}.
\label{c 7}
\eeq

Consider the transformation of the quadratic contractions
of the curvature tensors
\beq
\sqrt{-g}\,R^{2}_{\mu\nu\al\be}
&=&\sqrt{-{\bar g}}
\,R^{\mu\nu\al\be}R_{\mu\nu\al\be}
=\sqrt{-{\bar g}}\,e^{(D-4)\sigma}
\,\Big\{\,{\bar R}^2_{\mu\nu\al\be}
\nonumber
\\
&+&
2(D-2)\Big[\,2({\bar \na}_\mu{\bar \na}_\nu\si)^2
- 4({\bar \na}_\mu{\bar \na}_\nu\si)
({\bar \na}^\mu\si)({\bar \na}^\nu\si)
\nonumber
\\
&+&
4({\bar \na}^2 \si)({\bar \na}\si)^2+(D-1)({\bar \na}\si)^4
\Big]
\nonumber
\\
&+& 8{\bar R}^{\mu\nu}
({\bar \na}_\mu\si{\bar \na}_\nu\si
-{\bar \na}_\mu{\bar \na}_\nu\si)
-4{\bar R}({\bar \na}\si)^2
+4({\bar \na}^2 \si)^2\,\Big\} \,,
\label{111}
\eeq
where we multiplied the expression by $\sqrt{-g}$ for
convenience.
\beq
\sqrt{-g}\,R^2_{\mu\nu}
&=& \sqrt{-{\bar g}}\,e^{(D-4)\sigma}
\,\Big\{\,{\bar R}^2_{\mu\nu}-2{\bar R}(\Box \si)
+(3D-4)({\bar \na}^2 \si)^2
\nonumber
\\
&+& (D-2)\Big[\,2{\bar R}^{\mu\nu}
({\bar \na}_\mu\si{\bar \na}_\nu\si-{\bar \na}_\mu{\bar \na}_\nu\si)
-2{\bar R}({\bar \na}\si)^2
\nonumber
\\
&+& (D-1)(D-2)({\bar \na}\si)^4
-2(D-2)({\bar \na}_\mu{\bar \na}_\nu\si)({\bar \na}^\mu\si)
({\bar \na}^\nu\si)
\nonumber
\\
&+& (D-2)({\bar \na}_\mu{\bar \na}_\nu\si)^2
+2(2D-3)({\bar \na^2} \si)({\bar \na}\si)^2\,\Big]\,\Big\}
\label{112}
\eeq
and
\beq
\sqrt{-g}\,R^2
=\sqrt{-g}\,e^{(D-4)\sigma}
\,\Big[\,{\bar R}-2(D-1)({\bar \na}^ 2\si)
-(D-1)(D-2)({\bar \na} \sigma)^2\Big]^2\,.
\label{113}
\eeq
For the square of the Weyl tensor the transformation is
very simple
\beq
\sqrt{-g}\,C^{2}_{\mu\nu\al\be}
=\sqrt{-{\bar g}}\,e^{(D-4)\sigma}\,{\bar C}^2_{\mu\nu\al\be}\,.
\label{114}
\eeq

Another important combination of the quadratic in curvature
invariants is
\beq
E = R_{\mu\nu\al\be}R^{\mu\nu\al\be}
-4 \,R_{\al\be}R^{\al\be} + R^2\,.
\label{115}
\eeq
In $D=4$ it is the integrand of the Gauss-Bonnet topological
term (or Euler characteristic, see, e.g. \cite{DNF})
\beq
\int d^4x\, \sqrt{-g}\,E\,.
\label{116}
\eeq
But, even for the $D\neq 4$ the expression
does not contribute to the propagator of gravitons (traceless and
completely transverse modes of the metric perturbations
on the flat background). For this reason this term plays very
special role in string theory \cite{GSW}. The conformal
transformation of the Gauss-Bonnet integrand is
\beq
\sqrt{-g}E
&=&
\sqrt{-{\bar g}}e^{(D-4)\sigma}
\,\Big[\,{\bar E}
+ 8(D-3){\bar R}^{\mu\nu}({\bar \na}_\mu{\bar \na}_\nu\si
- {\bar \na}_\mu\si{\bar \na}_\nu\si)
\nonumber
\\
&-& 2(D-3)(D-4){\bar R}({\bar \na}\si)^2
+ 4(D-2)(D-3)^2({\bar \na}^2 \si)({\bar \na}\si)^2
\nonumber
\\
&-&  4(D-2)(D-3)({\bar \na}_\mu{\bar \na}_\nu\si)^2
+ 4(D-2)(D-3)({\bar \na}^2 \si)^2
\nonumber
\\
&+&  8(D-2)(D-3)({\bar \na}_\mu\si{\bar \na}_\nu\si)
({\bar \na}^\mu{\bar \na}^\nu\si)
-4(D-3)R({\bar \na}^2 \si)
\nonumber
\\
&+& (D-1)(D-2)(D-3)(D-4)({\bar \na}\si)^4
\,\Big] \,.
\label{117}
\eeq

The last remaining invariant is the surface term, which has the
following transformation rule:
\beq
\sqrt{-g}\,({\na}^2 R)
&=&
\sqrt{-{\bar g}}\,e^{(D-4)\sigma}\,\Big[\,{\bar \na}^2 {\bar R}
-2(D-4)R({\bar \na}\si)^2 - 2R{\bar \na}^2 \si
\nonumber
\\
&-& 2(D-1)(D-6)({\bar \na}^\mu\si)({\bar \na}_\mu\Box\si)
+ (D-6)({\bar \na}^\mu\si)({\bar \na}_\mu {\bar R})
\nonumber
\\
&+& 2(D-1)(D-2)(D-4)({\bar \na}\si)^4
+2(D-1)(3D-10)(\bar{\na}^2 \si)({\bar \na}\si)^2
\nonumber
\\
&-& 2(D-1)({\bar \na}^4 \si)
-(D-1)(D-2)(D-6)({\bar \na}^\mu\si){\bar \na}_\mu ({\bar \na}\si)^2
\nonumber
\\
&+&
4(D-1)(\Box \si)^2
-(D-1)(D-2)\Box ({\bar \na} \si)^2
\,\Big] \,.
\label{118}
\eeq

In deriving the last expression we used the transformation
of the operator ${\Box}$ acting on scalars
$$
{\na}^2
=e^{-2\sigma}\,[\,{\bar \na}^2
+(D-2)({\bar \na}^\mu \si){\bar \na}_\mu\,]\,.
$$

It is remarkable that in $D=4$ the following simple
relation takes place
\beq
\sqrt{-g}\,\Big(E-\frac23\,{\Box} R\Big)\,=\,
\sqrt{-\bar{g}}\,\Big({\bar E}-\frac23\,{\bar \na}^2 {\bar R}
+ 4{\bar \De_4}\si \Big)\,,
\label{119}
\eeq
where $\De_4$ is the fourth order Hermitian conformal invariant
operator acting on conformally invariant scalar field
\cite{pan,rei}
\beq
\De_4 = \na^4 + 2\,R^{\mu\nu}\na_\mu\na_\nu - \frac23\,R{\Box}
+ \frac13\,(\na^\mu R)\na_\mu\,.
\label{120}
\eeq
In section 4 we shall see that the formula (\ref{119})
is useful in understanding the relation between the
topological terms in $D=4$ and $D=2$.

\section{Practical use of conformal transformation}

Let us consider the practical application of the $D$-dimensional
conformal transformations derived in the previous section. We
shall also use the factorization theorem of section 2.

\subsection{Derivation of Einstein equations for the FRW metric}

Consider the derivation of the Einstein tensor for the
Friedmann-Robertson-Walker (FRW) metric in terms of the

conformal time $\,\eta$
\beq
ds^2 \,=\, g_{\mu\nu}\,dx^\mu \,dx^\nu
\,=\,a^2(\eta)\,\left(d\eta^2-dl^2\right)\,,
\label{FRW 1}
\eeq
where (see, e.g. \cite{Weinberg72})
\beq
dl^2\,=\,\frac{dr^2}{1-kr^2}+r^2d\th^2+r^2\sin^2 \th d\ph^2\,,
\label{FRW 2}
\eeq
where $\,k=0,\,\pm 1$.
In order to fit with our notations in section 3 we denote
\beq
a(\eta)=e^{\si(\eta)}\,,\qquad
g_{\mu\nu}\,=\,{\bar g}_{\mu\nu}\,e^{2\si}\,.
\label{FRW 3}
\eeq
It is easy to see that the new metric
\beq
{\bar g}_{\mu\nu}\,=\,\mbox{diag}\,\left(1,\,
-\frac{1}{1-kr^2},\,-r^2,\,-r^2\sin^2 \th \right)
\label{FRW 4}
\eeq
is factorized and therefore satisfies the conditions
of the factorization theorem. One of the consequences
is that the FRW metric is Weyl flat. This becomes
clear if we remember the transformation rule for the
Weyl tensor (\ref{c 7}) and also that
$\,C^\al\,_{\be\mu\nu}=0\,$ in $\,D=3\,$ dimensions
(see also \cite{tuk}).

The relation between $R_{\mu\nu}$ and
${\bar R}_{\mu\nu}$ is given by Eq. (\ref{c Ricci})
and the relation between $R$ and ${\bar R}$ by
eq. (\ref{c scalar}). In both cases one has to set
$D=4$. Consider, as an example, $R_{00}$.
Using (\ref{c Ricci}) we arrive at
$$
R_{00}={\bar R}_{00}-3\si^{\prime\prime}\,,
$$
where the prime indicates the derivative with
respect to $\eta$. Furthermore ${\bar R}_{00}=0$,
because it is a Ricci tensor of a one-dimensional
metric (remember factorization theorem). Therefore
\beq
R_{00}=-3\si^{\prime\prime}\,=\,-\,
\frac{3}{a^2}\,\Big(aa^{\prime\prime}-{a^{\prime}}^2\Big)\,.
\label{FRW 5}
\eeq

Further calculations involve ${\bar R}_{ij}$, where
$i,j=1,2,3$. ${\bar R}_{i0}=0$ automatically.
So, in fact we reduced the $D=4$ calculations
to the simpler $D=3$ ones. The components of ${\bar R}_{ij}$
can be easily calculated directly (this is technically
simpler) or through another
conformal transformation in $D=3$
\beq
{\bar g}_{ij}\,=\,\mbox{diag}\,
\left(-\frac{1}{1-kr^2},\,-r^2,\,-r^2\sin^2 \th \right)
\,=\,\ga_{ij}\,e^{2\rho(r)}\,,\qquad \rho(r)=\ln r.
\label{FRW 6}
\eeq
It is easy to see that the new metric is again factorized
and the calculation can be performed in a metric which is
a product of the one-dimensional metric $\ga_{11}\,$
and the two-dimensional metric
$\,\diag \big(\ga_{22},\,\ga_{33}\big)$. Let us
present only a final result ${\bar R}_{ij}=-2k\,{\bar g}_{ij}$.
After performing contraction, we obtain ${\bar R}=-6k$.
Now we are in a position to derive
\beq
R_{i0}=0 \qquad  \mbox{and} \qquad
R_{ij}\,=\,-\,{\bar g}_{ij}\left(2k + \si^{\prime\prime}
+ 2{\si^{\prime}}^2\right)\,.
\label{FRW 7}
\eeq

Then, after
applying Eq. (\ref{c scalar}) with $D=4$ we arrive at
\beq
R\,=\,-\,6\,e^{-2\si}\,\Big(k+\si^{\prime\prime}
+ {\si^{\prime}}^2\Big)
\,=\,-\,6\,\left(\frac{k}{a^2}
+ \frac{a^{\prime\prime}}{a^3} \right)\,.
\label{FRW 8}
\eeq
Finally, the result for the Einstein tensor has the
form
\beq
G_{\eta\eta} &=& R_{\eta\eta}-\frac12\,R\,g_{\eta\eta}
\,=\,3k + 3\left(\frac{a^\prime}{a}\right)^2\,,
\nonumber
\\
G_{\eta\,i}&=&0\,,
\nonumber
\\
G^i_j\,&=&\de^i_j\,\left( \frac{k}{a^2}
+\frac{ 2a^{\prime\prime} }{a^3}
-\frac{ {a^\prime}^2}{a^4}\right)\,.
\label{FRW 9}
\eeq
The transition to the physical time $t$ can be easily
performed through the tensor transformation rule
and the standard relation between the two time units
$dt=a(\eta)d\eta$
\beq
G_t^t &=&G_{tt}
= \left(\frac{d\eta}{dt}\right)^2\,G_{\eta\eta}
= \frac{3({\dot a}^2+k)}{a^2}\,.
\nonumber
\\
\nonumber
\\
G^i_j\,&=&\de^i_j\,\left( \frac{k}{a^2}
+\frac{ 2{\ddot a}}{a}
+ \frac{ {\dot a}^2}{a^2}\right)\,.
\label{FRW 10}
\eeq
Using the expressions (\ref{FRW 10}) one can
easily write down the standard form of the
Friedmann equations
\beq
\frac{ {\dot a}^2 + k }{a^2}\,=\,\frac{8\pi G\,\rho}{3}\,,
\qquad\qquad
 \frac{2{\ddot a}}{a}
+ \frac{ {\dot a}^2}{a^2} + \frac{k}{a^2}\,=\,-8\,\pi G\,P\,,
\label{FRW 11}
\eeq
where $P$ and $\rho$ are pressure and energy density of
the ideal fluid filling the Universe. These two
quantities are related
by the equation of state specific for the given type
of fluid (matter, radiation, vacuum energy etc).
From our point of view, the method of deriving the
Friedmann equations presented above is simpler than the
standard one. Of course, this is true only if one
has calculated the conformal transformations for the
Ricci tensor and scalar curvature first, but
these calculations are not involved.

\subsection{Derivation for the Schwarzschild metric}

The Schwarzschild metric is another important particular
example of the metrics used in GR. It is characterized by
spherical symmetry and therefore applies to the description
of the gravitational field in the exterior of numerous
objects like starts at the different stages of their
evolution. The extreme example is the point-like spherically
symmetric mass distribution which corresponds to the black
hole solution of GR. The formation and properties of the
black holes is extremely interesting subject (see, e.g.
the comprehensive monograph \cite{Frolov}). For this
reason, and also due to its relative simplicity, the
Schwarzschild solution is compulsory element of any course
of GR. The consideration of this solution always starts
with the derivation of Einstein equations for the corresponding
metric
\beq
ds^2&=&g_{\mu\nu}dx^\mu dx^\nu
\,=\,e^{\nu(r,t)}{dt^2}-e^{\la(r,t)}{dr^2}-r^2d\Om\,,
\label{b h 1}
\\
\mbox{where}\qquad
d\Om &=& \sin^2 \th d\ph^2 + d\th^2\,.
\nonumber
\eeq
The purpose of this section is the perform this calculation in
a more economic way\footnote{The method equivalent to the
one described below was used in the field theory textbook by W.
Siegel \cite{Siegel-1999}.} using the local conformal transformation
described in section 3. But, due to the relative simplicity
of this way of calculations, we will be able to consider
a little bit more complicated metric
\beq
ds^2 &=& g_{\mu\nu}dx^\mu dx^\nu
\,=\,e^{\nu(r,t)}{dt^2}-e^{\la(r,t)}{dr^2}-e^{2\Phi(r,t)}d\Om\,,
\label{b h 2}
\eeq
where $\,\Phi(r,t)\,$ is an arbitrary scalar function. Let us
notice that the more general metric (\ref{b h 2}) is widely
used in the black hole physics (see, e.g. \cite{Frolov}), in
particular because it has some advantages in the theories
with vacuum quantum corrections to the Einstein GR
\cite{balsan}. The particular case (\ref{b h 1}) may be
always achieved by replacing $\,\,\Phi(r,t)=\ln r\,\,\,$ into
the corresponding expressions.

Our strategy in calculating the Einstein equations for
the metric (\ref{b h 2}) will be the following. The first
step is the conformal transformation
\beq
g_{\mu\nu}\,=\,{\bar g}_{\mu\nu}\,e^{2\Phi}\,.
\label{b h 3}
\eeq
The new metric $\,{\bar g}_{\mu\nu}\,$ is diagonal and
factorized
$$
{\bar g}_{ab}
\,=\,\diag \left({\bar g}_{00} ,\,{\bar g}_{11}\right)
\,,\qquad
{\bar g}_{ij}
\,=\,\diag \left({\bar g}_{22} ,\,{\bar g}_{33}\right)\,,
$$
where
\beq
{\bar g}_{00}=e^{A}\,,\quad
{\bar g}_{11}=-e^{B}\,,\quad
{\bar g}_{22}= -1 \,,\quad
{\bar g}_{33}= -\sin^2\th \,,
\label{b h 4}
\eeq
and
\beq
A=A(r,t)=\nu(r,t)-2\Phi(r,t)
\,,\qquad
B=B(r,t)=\la(r,t)-2\Phi(r,t)\,.
\label{b h 5}
\eeq
The relations between the two Ricci tensors and the two
scalar curvatures are given by the particular form of
(\ref{c Ricci})
\beq
R_{\mu\nu}&=&{\bar R}_{\mu\nu}
- {\bar g}_{\mu\nu}({\bar \na}^2 \si)
+ 2({\bar \na}_\mu\si)({\bar \na}_\nu\si)
- 2({\bar \na}_\mu {\bar \na}_\nu\si)
- 2{\bar g}_{\mu\nu}({\bar \na}\si)^2\,,
\nonumber
\\
R&=&e^{-2\sigma}\,\Big[\,{\bar R}
- 6({\bar \na}^2 \sigma)
- 6({\bar \na} \sigma)^2\,\Big]\,.
\label{c scalar 4}
\eeq

According to the factorization theorem of section 2,
all components of the curvature tensor with the mixed
indices vanish and we need to calculate only the
curvature tensors and other quantities which emerge in
(\ref{c scalar 4}) for the two-dimensional metrics
in (\ref{b h 4}). For the curvatures we have
\beq
K_{abcd}=\frac12\,K\,
\left({\bar g}_{ac}\,{\bar g}_{bd}
-{\bar g}_{ad}\,{\bar g}_{bc}\right)\,,\qquad
a,b,... = t,r\,\,,
\label{b h 6}
\eeq
where $K_{abcd}$ is the Riemann tensor for the metric
$\,{\bar g}_{ab}\,$ and
\beq
k_{ijkl}=\frac12\,k\,
\left({\bar g}_{ik}\,{\bar g}_{jl}
-{\bar g}_{il}\,{\bar g}_{jk}\right)
\,,\qquad i,j,... = \th,\ph\,,
\label{b h 7}
\eeq
where $\,k\,$ is nothing but the scalar curvature of the
$D=2$ sphere $\,k=-2$, the negative sign appears due to
the negative sign of the metric components in (\ref{b h 4}).

For the metric ${\bar g}_{ab}$ we obtain, by direct calculation
\beq
{\bar \Ga}^t_{tt}=\frac{{\dot A}}{2}
\,,\quad
{\bar \Ga}^r_{tt}=\frac{A^\prime}{2}\,e^{A-B}
\,,\quad
{\bar \Ga}^t_{rt}=\frac{A^\prime}{2}
\nonumber
\\
{\bar \Ga}^r_{tr}=\frac{{\dot B}}{2}
\,,\quad
{\bar \Ga}^t_{rr}=\frac{{\dot B}}{2}\,e^{B-A}
\,,\quad
{\bar \Ga}^r_{rr}=\frac{B^\prime}{2}\,,
\label{b h 8}
\eeq
where the dot stands for $d/dt$ and the prime for the $d/dr$.

After a small algebra we obtain
\beq
K=\frac12\,e^{-A}\,\big({\dot A}{\dot B}-2{\ddot B} -{\dot B}^2\big)
+\frac12\,e^{-B}\,\big(2A^{\prime\prime}-A^\prime B^\prime
+ {A^\prime}^2\big)\,,
\label{b h 9}
\eeq
and check that
$\, K_{tt}=\frac{K}{2}\,e^A \,\, \mbox{and} \,\,
K_{rr}=-\frac{K}{2}\,e^B
\,\, \mbox{according to (\ref{b h 6})}$.
Furthermore
\beq
{\bar \na}_t{\bar \na}_t\Phi &=& {\ddot \Phi}
- \frac12\,{\dot A}{\dot \Phi}
- \frac12\,A^\prime \Phi^\prime\,e^{A-B}\,,
\nonumber
\\
{\bar \na}_r{\bar \na}_r\Phi &=& \Phi^{\prime\prime}
- \frac12\,B^\prime \Phi^\prime
- \frac12\,{\dot B}{\dot \Phi}\,e^{B-A}\,,
\nonumber
\\
{\bar \na}_t{\bar \na}_r\Phi &=& {\dot \Phi}^\prime
- \frac12\,A^\prime{\dot \Phi}
- \frac12\,{\dot B} \Phi^\prime\,.
\label{b h 10}
\eeq
and
\beq
\left({\bar \na \Phi}\right)^2
&=& e^{-A}\,{\dot \Phi}^2 \,-\, e^{-B}\,{\Phi^\prime}^2\,,
\nonumber
\\
{\bar \na}^2\Phi &=& e^{-A}\,\Big(
{\ddot \Phi}-\frac12\,{\dot A}{\dot \Phi} +
\frac12\,{\dot B}{\dot \Phi}\Big)
\,+\,
e^{-B}\,\Big(
\frac12\,B^\prime\Phi^\prime - \frac12\,A^\prime\Phi^\prime
-\Phi^{\prime\prime}\Big)\,.
\label{b h 11}
\eeq
Now we are in a position to calculate $R$ and $R_{\mu\nu}$.
Using the second equation (\ref{c scalar 4}), relation
$\,{\bar R}=K+2\,$ and formulas (\ref{b h 11}), we arrive at
\beq
R&=&-2e^{-2\Phi}+
e^{-\nu}
\Big[(\dot{\nu}-\dot{\la})\Big(2\dot{\Phi}
+\frac{\dot{\la}}{2}\Big)
-\ddot{\la}
-4\ddot{\Phi}-6\dot{\Phi}^2\Big]
\nonumber
\\
&+& e^{-\la}\Big[{\nu^{\prime \prime}}
+ \Big(2\Phi^\prime + \frac{\nu^{\prime}}{2}\Big)
(\nu^{\prime}-\la^{\prime})
+ 4\Phi^{\prime \prime}
+6\Phi^{\prime\,2}\Big]\,.
\label{b h 12}
\eeq
Similarly, using the first equation (\ref{c scalar 4})
we obtain the components of the Ricci tensor and finally
of the Einstein tensor. For example,
\beq
R_{\th\th}\,=\,-2e^{-2\Phi}+
e^{-\nu}
\Big[ \dot{\Phi}
\Big(\frac{\dot{\nu}}{2}-\frac{\dot{\la}}{2} -2\dot{\Phi}\Big)
-\ddot{\Phi}\Big]
+ e^{-\la}\Big[\Phi^{\prime \prime} +  \Phi^{\prime}
\Big(\frac{\nu^{\prime}}{2}-\frac{\la^{\prime}}{2}
+2\Phi^{\prime}\Big) \Big]
\label{b h 13}
\eeq
leads to
\beq
G^\th_\th
&=& R^\th_\th-\frac12\,R\,\de^\th_\th
\,=\, e^{-\nu}\,
\Big[\ddot{\Phi}+\ddot{\la}+\dot{\Phi}^2
+ \frac{\dot{\Phi}\,(\dot{\la}-\dot{\nu})}{2}
+ \frac{\dot{\la}\,\big(\dot{\la}-\dot{\nu}
+ 2\dot{\Phi}\big)}{4}\Big]
\nonumber
\\
&+& e^{-\la}\,\Big[
\frac{\Phi^{\prime}\,(\la^{\prime}-\nu^{\prime})}{2}
- \Phi^{\prime \prime} - {\Phi^{\prime}}^2
-\frac{\nu^{\prime \prime}}{2}
+\frac{\nu^{\prime}\,\big(\la^{\prime}-\nu^{\prime}\big)}{4}
\Big]\,.
\label{b h 14}
\eeq
In the special case $\,\Phi=\ln r\,$ (\ref{b h 14}) boils
down into
\beq
G^\th_\th
&=& e^{-\nu}\,
\Big(
  \frac{\dot{\la}^2}{4}
- \frac{\dot{\nu}\dot{\la}}{4}
+ \frac{\ddot{\la}}{2}\Big)
+ e^{-\la}\,\Big(\frac{\nu^\prime\la^\prime}{4}
-\frac{\nu^{\prime\, 2}}{4}
-\frac{\nu^{\prime\prime}}{2}
+ \frac{\la^{\prime}-\nu^{\prime} }{r}\Big)\,,
\label{b h 15}
\eeq
in a perfect fit with the well-known result \cite{LL-2}.

In a similar way, using (\ref{c scalar 4}) and
(\ref{b h 10}) we obtain
 \beq
G_t^r &=& g^{rr}G_{tr}\,=\,g^{rr}R_{tr}-\frac{1}{2}\de_t^r\,R
\,=\,g^{rr}R_{tr}
\nonumber
\\
 &=&
2e^{-\la}\,\Big(\dot{\Phi}^{\prime}-\frac{\nu^{\prime}\dot{\Phi}}{2}
-\frac{\dot{\la}\Phi^{\prime}}{2}+\Phi^{\prime}\dot{\Phi}\Big)\,,
\label{b h 16}
\eeq
\beq
G_r^r &=& g^{rr}R_{rr} - \frac{1}{2}\de_r^r\,R
\,=\,-e^{-\la}R_r^r-\frac{1}{2}\,R
\nonumber
\\
 &=& e^{-2\Phi}-e^{-\la}
\left(\nu^{\prime}\Phi^{\prime}+\Phi^{\prime\,2}\right)
\,+\,
e^{-\nu}\left(2\ddot{\Phi}+3\dot{\Phi}^2-\dot{\nu}\dot{\Phi}\right)
\,,
\label{b h 17}
\eeq
\beq
G_t^t &=& g^{tt}R_{tt}-\frac{1}{2}\,R
\nonumber
\\
 &=& e^{-2\Phi}+e^{-\nu}\,\Big(\dot{\la}\dot{\Phi}+\dot{\Phi}^2)
\,+\,e^{-\la}\,[-2\Phi^{\prime \prime}+\la^{\prime}\Phi^{\prime}
-3\Phi^{\prime\,2}\Big)\,.
\label{b h 18}
\eeq
In the special case $\,\Phi=\ln r\,$ these expressions become
the standard ones \cite{LL-2}
\beq
G_t^r &=& -e^{-\la}\dot{\la}\ph^{\prime}
\,=\,-\frac{\dot{\la}}{r}e^{-\la}\,,
\nonumber
\\
\nonumber
\\
G_r^r &=& -e^{-\la}\,\Big(\frac{\nu^{\prime}}{r}
+\frac{1}{r^2}\Big)+\frac{1}{r^2}\,,
\nonumber
\\
\nonumber
\\
G_t^t &=& e^{-\la}\Big(\frac{\la^{\prime}}{r}
-\frac{1}{r^2}\Big)+\frac{1}{r^2}\,.
\label{b h 19}
\eeq
Let us remark that if the derivation is performed
directly for the usual metric (\ref{b h 1}), it is
technically very simple and definitely less cumbersome
than the standard one without conformal transformations.

\section{Dimensional reduction of the topological invariant}

The purpose of this section is to illustrate the
calculational power of the conformal transformation
for the theories beyond the GR. Let us use this
transformation for deriving the form of the $\,D=4\,$
Gauss-Bonnet
topological invariant (\ref{116}) in the particular
metric
\beq
g_{\mu\nu} \,=\,\left\| \begin{array}{cc}
 g_{ab}(x^a) & 0 \\
0 & e^{2\Phi(x^a)}h_{ij}(x^i) \end{array}\right\|\,.
\label{g b 0}
\eeq
Here $g_{ab}(x^a)$ and $h_{ij}(x^i)$ are both
two-dimensional metrics, but $h_{ij}(x^i)$ has constant scalar 
curvature $\,\,k\,\,$ while $g_{ab}(x^a)$ is arbitrary. $\Phi(x^a)$ 
is a scalar field in the two-dimensional space. The result of this
dimensional reduction is supposed to be the topological or surface 
term in the two-dimensional space with the metric $g_{ab}$,
$\,a,b=0,1$. Taking into account the simplicity of the 
conformal transformation for the combination (\ref{119}), we shall 
consider not the proper Gauss-Bonnet term, but instead its linear
combination with the surface term
\beq
S_{top}\,=\,\int d^4x\sqrt{-g}
\,\left(E - \frac23\,\na^2 R\right)\,.
\label{g b 1}
\eeq
After the conformal transformation
$\,g_{\mu\nu}={\bar g}_{\mu\nu}\cdot\exp (2\Phi)$,
according to (\ref{119}), we obtain
\beq
S_{top}\,=\,\int d^4x
\sqrt{-{\bar g}}\left({\bar E}
- \frac23\,{\bar \na}^2 {\bar R} + 4{\bar \De}_4\Phi\right)\,.
\label{g b 2}
\eeq
Since the metric ${\bar g}_{\mu\nu}$ is factorized
\beq
{\bar g}_{\mu\nu} \,=\,\left\| \begin{array}{cc}
 \bar{g}_{ab}(x^a) & 0 \\
       0 & h_{ij}(x^i) \end{array}\right\|\,,
\label{g b 3}
\eeq
we need to calculate all the relevant (and complicated)
curvature-squared expressions
only for the two-dimensional metrics $\,g_{ab}(x^a)\,$ and
$\,h_{ij}$. The quantities based on the metric of the
two-dimensional sphere $\,h_{ij}\,$ are very simple
due to Eq. (\ref{number 3}), and the whole problem can
be resolved immediately. Let us denote the curvature tensor
corresponding to the metric $\,g_{ab}\,$ by
$\,{\cal K}_{abcd}\,$
and the curvature tensor corresponding to the metric
$\,{\bar g}_{ab}\,$ by $\,{\bar {\cal K}}_{abcd}$. Indeed,
as always in $\,D=2$, we have (\ref{number 3}), that is
\beq
{\bar {\cal K}}_{abcd} &=& \frac12\,{\bar {\cal K}}
\,\big( {\bar g}_{ac}{\bar g}_{bd}-
{\bar g}_{ad}{\bar g}_{bc}\big)\,,\qquad
{\bar {\cal K}}_{ab} = \frac12\,{\bar {\cal K}}\,{\bar g}_{ab}
\nonumber
\\
{\cal K}_{abcd} &=& \frac12\,{\cal K}
\,\big( g_{ac}g_{bd} - g_{ad}g_{bc}\big)\,,\qquad
{\cal K}_{ab} = \frac12\,{\cal K}\,g_{ab}\,.
\label{g b 4}
\eeq
Furthermore, in the $h_{ij}$ sector the Riemann and Ricci
tensors are
$$
k_{ijkl} = \frac12\,k
\,\big( h_{ik}h_{jl} - h_{il}h_{jk}\big)\,,\qquad
k_{ij} = \frac12\,k\,h_{ij}\,.
$$
Let us calculate the elements of the expression
(\ref{g b 2}).
\beq
{\bar E}\,=\,{\bar {\cal K}}_{abcd}{\bar {\cal K}}^{abcd}
- 4 {\bar {\cal K}}_{ab}{\bar {\cal K}}^{ab}
+ k_{ijkl}k^{ijkl}-4k_{ij}k^{ij}
+ \left({\bar {\cal K}} + k\right)^2 \,=\,2k\bar{\cal K}
\label{g b 6}
\eeq
The covariant derivatives corresponding to the metrics
$\,g_{ab}\,$ and $\,{\bar g}_{ab}\,$ will be denoted by
$\,{\cal D}_a\,$ and $\,{\bar {\cal D}}_a\,$ correspondingly.
When acting on scalar (e.g. $\Psi$), the transformation
rule for the d$^\prime$Alembert operator looks like
\beq
{\cal D}^2\Psi\,=\,e^{-2\Phi}{\bar {\cal D}}^2\Psi\,.
\label{g b 7}
\eeq

For the last term in Eq.~(\ref{g b 2}), it proves useful to rewrite
the operator (\ref{120}) in another form, that shows explicitly
that it is a derivative operator,
\beq
\bar{\De}_4 \,=\,
\bar{\na}^4 + 2 \bar{\na}_\mu \bar{R}^{\mu\nu}\bar{\na}_\nu
- \frac23\,  \bar{\na}^\mu \bar{R} \bar{\na}_\mu,
\label{dera}
\eeq
where we use the notations $\na A = A \na + (\na A)$.

Using (\ref{g b 4}), (\ref{c scalar}), (\ref{g b 7}) and
(\ref{dera}), we obtain
\beq
\bar{\De}_4 \Phi
&=&
{\bar{\cal D}}^4\Phi
+ \bar{\cal D}^a {\bar {\cal K}}\,{\bar {\cal D}}_ a\Phi
- \frac23\,
 \bar{\cal D}^a (k + {\bar {\cal K}})\,{\bar {\cal D}}_ a\Phi
\nonumber
\\
&=&
e^{2\Phi}\left[
{\cal D}^2 \,e^{2\Phi}\,{\cal D}^2 \Phi
\,+\, \frac13\,{\cal D}^a e^{2\Phi}
\left({\cal K} + 2{\cal D}^2\Phi\right){\cal D}_a \Phi
\,-\, \frac83\,k {\cal D}^2 \Phi\Big)\right]
\label{g b 8}
\eeq
and
\beq
{\bar \na}^2 {\bar R}\,=\,e^{2\Phi}\,{\cal D}^2
\left[ e^{2\Phi}\, \left({\cal K}
+ 2{\cal D}^2\Phi\right)\right]\,.
\label{g b 9}
\eeq
Finally, for the modified topological term (\ref{g b 2}) we find,
adding the overall factor of the volume $V_2$ of the space with
the the metric $h_{ij}$,
\beq
S_{top}
&=&
V_2 \int d^2x
\sqrt{-g(x^a)} \,\bigg\{\,2k\,{\cal K}
\,+\,\frac43\,k {\cal D}^2 \Phi
\,-\, \frac23\,{\cal D}^2 \big(e^{2\Phi}{\cal K})
\nonumber
\\
&&
\,+\,\frac83\, {\cal D}^2 \big(e^{2\Phi}{\cal D}^2\Phi)
+ \frac43\,{\cal D}^a \Big[e^{2\Phi} \big({\cal K})
+ 2{\cal D}^2\Phi\big){\cal D}_a\Phi\Big]\bigg\}\,.
\label{g b 10}
\eeq
The first term here is the Einstein-Hilbert action
(with the coefficient $2k$) which is indeed topological
term in $D=2$. Other terms are surface integrals which
depend on the boundary conditions only. Therefore the
expression (\ref{g b 10}) is a kind of a direct relation
between the topological invariants in $D=4$ and $D=2$.

If we consider similar dimensional reduction to $D=3$
space, the topological invariant does not emerge, and
we meet only surface integrals. In order to see
this one does not need to perform calculations. It is
enough to look at (\ref{g b 2}) and remind that the
potential source of the topological terms ${\bar E}$
vanish in $D=3$, because it coincides with the square
of the Weyl term $\,{\bar C}^2_{abcd}$.

\section{Conclusions}

We have pedagogically reviewed the local conformal
transformations and demonstrated their utility in
relatively easy derivation of various quantities
related to the gravitational theory. In particular,
using these transformations and the factorization
theorem of section 2 one can essentially reduce the
amount of work needed to obtain the Einstein
equations for the FRW and Schwarzschild metrics. From
our point of view, this way of calculation provides
certain advantages compared to the direct one, and
can be used in the basic courses of GR.

Furthermore, we applied the same scheme of calculation
to the fourth order terms in the gravitational
action. These terms emerge frequently as quantum
corrections to GR, in particular in the frameworks
of semiclassical approach to gravity and in string
theory. The application of conformal
transformations made the calculations fairly easy, while
the direct way of deriving the same quantities would
require an enormous amount of work. As a result
of this calculus we have obtained a direct relation
between the topological invariants in $D=4$ and $D=2$
spaces. In $D=3$ case the same procedure reduces the
Gauss-Bonnet term to the integral of total derivative
and there is no relation to the corresponding
topological (Chern-Simons) action.

\section*{Acknowledgements}

Authors are grateful to CNPq for research fellowship (I.L.Sh.)
and also to CNPq (B.G.), FAPEMIG (D.F.) and UFJF (A.de L.)
for the scientific initiation scholarship.

\section*{Acknowledgements to v2}

I.Sh. is very grateful to Leslaw  Rachwa\l,  \ Nobuyoshi Ohta and 
V\'{\i}tor Fernandes Barra for pointing out to me a number of 
misprints and mistakes in the formulas of the first version. 




\begin{thebibliography}{99}


\bibitem{LL-2}
L.D. Landau and E.M. Lifshits, ``Field Theory'',
Elsevier, 1995.

\bibitem{Weinberg72}
S. Weinberg, ``Gravitation and Cosmology'',
Wiley, New York, 1972.

\bibitem{MTW} C.W.Misner, K.S.Thorn and J.A.Wheeler,
``Gravitation'', Freeman, San Francisco, 1973.

\bibitem{wald} R.M. Wald, ``General Relativity'',
University of Chicago Press, Chicago, 1984.

\bibitem{birdav}
N.D. Birrell and P.C.W. Davies, ``Quantum fields in
curved space'', Cambridge University Press, Cambridge, 1982.

\bibitem{fulling} S.A. Fulling,
``Aspects of Quantum Field Theory in Curved Space-Time''
Cambridge Univ. Press, Cambridge, 1989.

\bibitem{book}
I.L. Buchbinder, S.D. Odintsov and I.L. Shapiro,
``Effective Action in Quantum Gravity'',
IOP Publishing, Bristol, 1992.

\bibitem{petrov}
A.Z. Petrov, ``Einstein Spaces'', Pergamon, Oxford, 1969.

\bibitem{DNF}
B.A. Dubrovin, A.T. Fomenko and S.P. Novikov,

``Modern Geometry-Methods and Applications'',
Springer-Verlag, 1984.

\bibitem{GSW}
M.B. Green, J.H. Schwarz and E. Witten,
``Superstring Theory'',
Cambridge University Press, Cambridge, 1987.

\bibitem{pan}
S. Paneitz,
``A Quartic Conformally Covariant Differential
Operator for Arbitrary Pseudo-Riemannian Manifolds'',
MIT preprint, 1983 (unpublished).

\bibitem{rei}
R.J. Reigert,
{\it Phys. Lett.} {\bf 134B} 56 (1980).

\bibitem{tuk}
A.P. Lightman, W. H. Press, R. H. Price and
S. A. Teukolsky, ``Problem Book in relativity and Gravitation'',
Princeton University Press, Princeton, 1975.

\bibitem{Siegel-1999} W.~Siegel,
\textit{Fields,}
(Free on-line textbook, hep-th/9912205).

\bibitem{Frolov}
V.P. Frolov and I.D. Novikov,
``Black Hole Physics - Basic Concepts and New Developments'',
Kluwer Academic Publishers, 1989.

\bibitem{balsan}
R. Balbinot, A. Fabbri and I.L. Shapiro,
{\it Phys. Rev. Lett.} {\bf 83} (1999) 1494;
{\it Nucl. Phys.} {\bf B559} (1999) 301.

\end{thebibliography}
\end{document}